\documentclass[10pt,conference]{IEEEtran}

\usepackage{xcolor,colortbl}
\usepackage{multirow}
\usepackage{graphicx}
\usepackage{arydshln}
\usepackage{amsmath}
\usepackage{flushend} 

\usepackage{tikz}
\usepackage{pgfplots}
\usetikzlibrary{plotmarks}

\usepackage{listings} 
\usepackage{color}
\usepackage{pifont}
\usepackage{url}

\usepackage[T1]{fontenc}
\usepackage[utf8]{inputenc}

\definecolor{dkgreen}{rgb}{0,0.6,0}
\definecolor{gray}{rgb}{0.5,0.5,0.5}
\definecolor{mauve}{rgb}{0.58,0,0.82}
\definecolor{Gray1}{gray}{0.95}
\definecolor{Gray2}{gray}{0.90}

\begin{document}

\IEEEoverridecommandlockouts
\title{EvoSpex: An Evolutionary Algorithm for Learning Postconditions}

\author{
\IEEEauthorblockN{Facundo Molina\IEEEauthorrefmark{1}\IEEEauthorrefmark{2}, Pablo Ponzio\IEEEauthorrefmark{1}\IEEEauthorrefmark{2}, Nazareno Aguirre\IEEEauthorrefmark{1}\IEEEauthorrefmark{2}, Marcelo Frias\IEEEauthorrefmark{2}\IEEEauthorrefmark{3}}
\IEEEauthorblockA{\IEEEauthorrefmark{1}Department of Computer Science, FCEFQyN, University of R\'io Cuarto, Argentina\\ 
}
\IEEEauthorblockA{\IEEEauthorrefmark{2}National Council for Scientific and Technical Research (CONICET), Argentina}
\IEEEauthorblockA{\IEEEauthorrefmark{3}Department of Software Engineering, Buenos Aires Institute of Technology, Argentina}
}
 
\maketitle

\begin{abstract}
Software reliability is a primary concern in the construction of software, and thus a fundamental component in the definition of software quality. Analyzing software reliability requires a \emph{specification} of the intended behavior of the software under analysis, and at the source code level, such specifications typically take the form of \emph{assertions}. Unfortunately, software many times lacks such specifications, or only provides them for scenario-specific behaviors, as assertions accompanying tests. This issue seriously diminishes the analyzability of software with respect to its reliability. 

In this paper, we tackle this problem by proposing a technique that, given a Java method, automatically produces a specification of the method's current behavior, in the form of postcondition assertions. This mechanism is based on generating executions of the method under analysis to obtain \emph{valid} pre/post state pairs, mutating these pairs to obtain (allegedly) \emph{invalid} ones, and then using a genetic algorithm to produce an assertion that is satisfied by the valid pre/post pairs, while leaving out the invalid ones. The technique, which targets in particular methods of reference-based class implementations, is assessed on a benchmark of open source Java projects, showing that our genetic algorithm is able to generate post-conditions that are stronger and more accurate, than those generated by related automated approaches, as evaluated by an automated oracle assessment tool. Moreover, our technique is also able to infer an important part of manually written rich postconditions in verified classes, and reproduce contracts for methods whose class implementations were automatically synthesized from specifications.
\end{abstract}

%


\section{Introduction}

The quality of software systems is typically defined around various dimensions, such as reliability, usability, efficiency, etc. Among these, reliability is in general considered a fundamental attribute of software quality, and a primary concern in software development \cite{Ghezzi+2002,DBLP:series/txcs/Jalote05}. Analyzing software reliability is strongly related to finding software defects, i.e., actual software behaviors that diverge from the expected behavior. Discovering such defects requires one to state \emph{what} the expected behavior is, in other words, a \emph{specification} of the software. Many times such specifications are either implicit, or stated informally, diminishing the possibility of exploiting specifications for (automated) reliability analysis. 

Software specifications can appear in different forms. At the level of source code, when present, they generally manifest either as \emph{comments}, i.e., informal descriptions of what the software is supposed to do, or more formally as \emph{program assertions}, i.e., (usually executable) statements that assert properties that the software must satisfy at certain points during program executions. The former are more common, but cannot be straightforwardly used for automated reliability analysis. The latter, on the other hand, are readily usable for program analysis, especially when stated as \emph{contracts} \cite{DBLP:journals/computer/Meyer92}, but they are seldom found accompanying source code. Moreover, many times program assertions state scenario-specific properties, e.g., statements that only express the expected software behavior for a test case, as opposed to the more general, and also significantly more useful, assertions associated with contract elements such as invariants and pre/post-conditions. 

Due to the above described situation regarding specifications at the level of source code, the \emph{specification inference problem} (a special case of the well known \emph{oracle problem} \cite{Barr+2015}), i.e., taking a program \emph{without} a corresponding specification and attempting to automatically produce one that captures the program's current behavior, is receiving increasing attention by the software engineering community. Automatically inferring specifications from source code is a relevant topic, as it enables a number of applications, including program comprehension, software evolution and maintenance, bug finding \cite{DBLP:conf/oopsla/SchillerE12}, and specification improvement \cite{DBLP:conf/oopsla/SchillerE12,DBLP:conf/issta/JahangirovaCHT16}, among others. 

In this paper, we tackle this problem by proposing a technique that, given a Java method, automatically produces a specification of the method's current behavior, in the form of postcondition assertions. This mechanism is based on generating \emph{valid} and \emph{invalid} pre/post state pairs (i.e., state pairs that represent, and do not represent, the method's current behavior, respectively), which guide a genetic algorithm to produce a JML-like assertion characterizing the valid pre/post pairs, while leaving out the invalid ones. The generation of valid pre/post pairs is based on executing the method on a \emph{bounded exhaustive} test set, generated by exercising the method inputs' APIs using user-defined ranges for basic datatypes, and bounding their execution sequences. The invalid pre/post pairs, on the other hand, are obtained by \emph{mutating} valid pairs, i.e., arbitrarily modifying the post-states so that each resulting pair does not belong to the set of valid pairs. This mutation-based approach to generate invalid pairs is unsound, in the sense that it may lead to \emph{valid} pairs instead, an issue that may affect the precision of the produced assertions. As we describe later on, the design of our genetic algorithm takes it into account. Because of the assertion language we consider, that involves quantification, object navigation and reachability expressions, our approach is particularly well-suited for reference-based class implementations with (implicit) strong representation invariants, such as heap-allocated structural objects, and complex custom types. 

We assess our technique on a benchmark of open source Java projects taken from \cite{DBLP:journals/tosem/FraserA14}, featuring complex implementations of reference-based classes. In these case studies, our genetic algorithm is able to generate post-conditions that are stronger and more accurate, than those generated by related specification-inference approaches, as evaluated by OASIs, an automated oracle assessment tool \cite{DBLP:conf/issta/JahangirovaCHT16}. Moreover, our technique is also able to infer an important part of manually written rich postconditions (strong contracts used for verification) present in verified classes \cite{TFNP-TACAS15}, and reproduce contracts for methods whose class implementations were automatically synthesized from specifications \cite{DBLP:conf/icse/LoncaricET18}. 


\section{Background}
\label{background}

\subsection{Assertions as Program Specifications}

The use of assertions as program specifications dates back to the works of Hoare \cite{DBLP:journals/cacm/Hoare69} and Floyd \cite{Floyd67}, in the context of program verification and associated with the concept of program correctness. Technically, an assertion is a statement predicating on program states, that can be used to capture \emph{assumed properties}, as in the case of preconditions, or \emph{intended properties}, as in the case of postconditions. A program $P$ accompanied by a precondition \emph{pre} and postcondition \emph{post} is said to be (partially) correct with respect to this specification, if every execution of $P$ starting in a state that satisfies \emph{pre}, if it terminates, it does so in a state that satisfies \emph{post} \cite{DBLP:journals/cacm/Hoare69}. That is, every valid terminating execution of $P$, i.e., every execution satisfying the requirements stated in the precondition, must lead to a state satisfying the postcondition. 

While program assertions originated in the context of program verification, they soon permeated into programming languages constructs and (informal) programming methodologies. More recently, they have been central to the definition of methodologies for software design, notably \emph{design by contract} \cite{Meyer1997}. Most modern imperative and object-oriented programming languages support assertions, either as built-in constructions \cite{Meyer1998} or through mature libraries such as Code Contracts \cite{DBLP:conf/rv/Barnett10} and JML \cite{DBLP:conf/fmco/ChalinKLP05}. Moreover, libraries for unit testing make extensive use of assertions to automate checking the expected results of running a test case. 

Preconditions are more commonly seen in source code, e.g., within methods in the form of state and argument checks, throwing appropriate exceptions when these are found invalid, preventing normal execution. Postconditions, on the other hand, are less common. Post-execution checks are commonly seen as part of test cases, although they rarely capture postconditions, in the sense of general properties that every execution must satisfy on termination; post-execution checks in tests generally state properties that should be satisfied for the specific test where they are stated.    

The assertion language that we consider in this paper is, from an expressiveness point of view, a JML-like \cite{DBLP:conf/fmco/ChalinKLP05} contract language. More precisely, we follow the approach used in \cite{Khalek+2011}, and use the Alloy notation \cite{Jackson2006}. The language supports quantifiers, navigation and reachability expressions including navigations through one or more field. A sample specification, generated by our technique, is shown in Figure~\ref{evospex-post-add-treelist}. Most operators have a direct intuitive reading (equality and inequalities, boolean connectives, etc.); \texttt{all} and \texttt{some} are the universal and existential quantifiers, respectively; the dot operator (\texttt{.}) is relational composition and captures navigation; relational union and intersection are denoted by \texttt{+} and \texttt{\&}, respectively, and can be applied to combine fields in navigations; set/relational cardinality is denoted by \texttt{\#}; finally, \texttt{*} and \texttt{\^{}} are reflexive-transitive closure and transitive closure, respectively. Closures allow the language to express reachability. For instance, the last sentence in Figure~\ref{evospex-post-add-treelist} expresses that for every node \texttt{n} reachable (in zero or more steps) from the root by traversing \texttt{left} and \texttt{right} (i.e., all nodes in the tree), it is not the case that \texttt{n} is included in the set of nodes reachable in one or more steps from \texttt{n} itself. That is, the \texttt{left + right} structure from the \texttt{root} is acyclic. It is worth to mention that all assertions in this language can be checked at run-time, and thus we can use it to assert properties in program points. We refer the reader to \cite{Jackson2006} for further details regarding the language. 

\subsection{Quality of Assertions}

As described above, program assertions are a way of capturing the expected software behavior via expressions that convey intended properties of program states in specific parts of a program. Such expected behavior can be captured with different degrees of precision, leading to assertions of different quality. The most typical issue with program assertions is the misclassification of invalid program states as valid. This is essentially the effect of having \emph{weak} assertions, that are able to detect some, but not all, faulty situations. It is rarely considered a \emph{defect} in the assertion, but an inherent issue associated with a balance between expressiveness and economy/efficiency in the definition of assertions. Indeed, it is even considered methodologically correct to express weak (and efficiently checkable) assertions \cite{Meyer1997}. Following the terminology put forward in \cite{DBLP:conf/issta/JahangirovaCHT16}, a real program execution leading to an invalid program state that a corresponding assertion is unable to detect is called a \emph{false negative}. 

A second issue with program assertions is the dual of the previous, i.e., the misclassification of valid program states as invalid. This issue indicates that the assertion is \emph{wrong}, as it does not properly specify the intended behavior of the software. Such issues are typically considered to be specification defects. This situation can also often arise as a consequence of software evolution, when required changes in program behavior are (correctly) implemented, but the accompanying assertions are not kept in synchrony with the evolved behavior \cite{DBLP:conf/kbse/DanielJDM09}. A real program execution leading to a valid program state, that a corresponding assertion classifies as an assertion violation, is a \emph{false positive}, according to the terminology put forward in \cite{DBLP:conf/issta/JahangirovaCHT16}.

Assessing the quality of assertions accompanying a program is a very challenging problem, that is typically performed manually. A way of measuring the quality of assertions is by attempting to determine the number of false positives and false negatives that a given assertion has. This idea has been exploited in \cite{DBLP:conf/issta/JahangirovaCHT16}, where an automated mechanism for evaluating the quality of assertions, based on evolutionary computation, is proposed. The approach presented therein executes an evolutionary test generation tool (the well-known tool EvoSuite \cite{DBLP:conf/sigsoft/FraserA11}) that tries to find false positives and false negatives, and when found, produces witnessing test cases, that can be used to (manually) improve the corresponding assertions. It is worth remarking that, for contracts specified in standard assertion languages, 
it is hardly expected for a contract to \emph{fully} capture the behavior of a program. As explained in \cite{DBLP:conf/vstte/PolikarpovaFM10}, precisely capturing a program's intended semantics requires additional mechanisms, such as the use of model classes, that imply the manual definition of abstractions of the state space of the program being specified. In terms of the above-mentioned issues with program assertions, it means that, technically, one can very often come up with false negatives, i.e., finding states that satisfy a given assertion but correspond to incorrect program behavior.

\section{An Illustrating Example}
\label{motivating-example}

As an illustrating example, let us consider a Java class implementing lists, partially shown in Figure~\ref{treelist}\footnote{This implementation is taken from https://www.nayuki.io/page/avl-tree-list}. This class implements list operations over balanced trees, supporting insertion and deletion from the list in $O(\log n)$, as opposed to the classic array-based and linked-list based list implementations. Let us focus on method \texttt{add}, that inserts an element in the list. Notice how the precondition of the method is captured in the source code, checking the validity of the index for insertion and that the tree has not reached its maximum size. The method postcondition, on the other hand, is not present in this implementation. Having the postcondition has multiple applications, in particular as assertions for testing future improvements of this method, and as a declarative description of what this method does (how it operates on the data structure), among many others. Writing the specification is, however, nontrivial, and thus coming up with the right expression for the postcondition is an important problem. 

\begin{figure}[t!]
\begin{lstlisting}
import java.util.AbstractList;

public final class AvlTreeList<E> extends AbstractList<E> {

	private Node<E> root; 

	public void add(int index, E val) {
		if (index < 0 || index > size())
			throw new IndexOutOfBoundsException();
		if (size() == Integer.MAX_VALUE)
			throw new IllegalStateException("Max size reached");
		root = root.insertAt(index, val);
	}

	private static final class Node<E> {

		private E value;
		private int height;
		private int size;
		private Node<E> left;
		private Node<E> right;

		public Node<E> insertAt(int index, E obj) {
			assert 0 <= index && index <= size;
			if (this == EMPTY_LEAF)
				return new Node<>(obj);
			int leftSize = left.size;
			if (index <= leftSize)
				left = left.insertAt(index, obj);
			else
				right = right.insertAt(index-leftSize-1, obj);
			recalculate();
			return balance();
		}
  
  }

}
\end{lstlisting}
\caption{Add method of class AvlTreeList}
\label{treelist}
\end{figure}

A well-known tool to assist the developer in this situation is Daikon \cite{Ernst+2007}. Daikon performs run-time invariant detection, it runs the program on a set of test cases, and observes which properties hold during these runs at particular program points, such as after method invokations. It then suggests as likely invariants those properties that were not falsified by any execution, or equivalently, that held true for all observed executions. The quality of the obtained invariants strongly depends on the program executions considered by Daikon (i.e., the set of tests that the user provides), and the set of candidate expressions to be considered. In particular for method \texttt{add} in Figure~\ref{treelist}, Daikon produces the postcondition shown in Figure~\ref{daikon-post-add-treelist}, when fed with \emph{all} valid tree lists of size up to 4. The shown postcondition is actually that produced by Daikon, but manually filtering out invalid expressions (inducing false positives) that could not be falsified by the suite. Still, as it can be seen, the postcondition generated in this case is relatively weak: we would expect to have some further information about how node attributes get manipulated in this implementation of lists over trees. The reason why Daikon produces this very simple postcondition in this case has to do with the set of candidate expressions that Daikon considers, which are produced from the definition of the program, and are restricted to relatively simple program properties (e.g., structural constraints, membership checking, etc., are not considered) \cite{Ernst+2007}.  

\begin{figure}[t!]
\begin{lstlisting}
// height
this.root.height >= old_this.root.height &&
this.root.height >= old_this.root.left.height &&
this.root.height >= old_this.root.right.height &&
// size
this.root.size > old_this.root.height &&
this.root.size > old_this.root.left.height &&
this.root.size > old_this.root.right.height &&
// left height
this.root.left.height <= old_this.root.height &&
this.root.left.height >= old_this.root.left.height &&
this.root.left.height >= old_this.root.right.height &&
// right height
this.root.right.height <= old_this.root.height && 
this.root.right.height >= old_this.root.left.height &&
this.root.right.height >= old_this.root.right.height  
\end{lstlisting}
\caption{Postcondition generated by Daikon for AvlTreeList.add(int, E)}
\label{daikon-post-add-treelist}
\end{figure}

Our aim is to provide stronger postconditions in cases such as the above. Our approach is, in essence, similar to Daikon's: the method under analysis is run for different inputs, and from information extracted from these runs we propose a postcondition for the method. There are, however, multiple differences. Firstly, our approach is based on generating runs for the method under analysis \emph{bounded exhaustively}, as opposed to Daikon, which requires tests to be provided (in the above example, the suite we provided Daikon with was the one that our technique produces). Our technique for generating the bounded exhaustive test suite is based on exercising the API of the inputs of the program under analysis, contrary to related approaches that require a specification \cite{Boyapati+2002,Khalek+2011}. Secondly, we consider both \emph{valid} and \emph{invalid} program states (although, as we explain later on, the approach to generate invalid states may unsoundly generate valid ones) in attempting to determine a method's postcondition, instead of only \emph{valid} executions, as is the case with Daikon. Third, our approach is based on evolving specifications, instead of considering non-falsified candidate properties. The details of our technique are described in the next section. Let us just mention that, for method \texttt{add} of class \texttt{AvlTreeList}, our obtained postcondition is the one shown in Figure~\ref{evospex-post-add-treelist}. Notice how the size update (referring to the relation between the pre and post states) and the membership of the inserted element are captured, as well as some structural properties of the representation.

\begin{figure}[t!]
\begin{lstlisting}
// root
this.root != null &&
this.root.left != null && 
// height
all n : this.root.*(left+right) : (
	n.left != null => n.height > n.left.height && 
	n.right != null => n.height > n.right.height 
) &&
// size
old_this.root.size < this.root.size &&
this.root.size == #(this.root.*(left+right - null)) - 1 &&
all n : this.root.*(left+right) : (
	n.left != null => n.size > n.left.size && 
	n.right != null => n.size > n.right.size 
) &&
// arguments
index != this.root.size &&
val in this.root.*(left+right).value &&
// structural
all n : this.root.*(left+right) : n !in n.^(left + right) 
\end{lstlisting}
\caption{Postcondition generated by our tool for AvlTreeList.add(int, E)}
\label{evospex-post-add-treelist}
\end{figure}










\section{EvoSpex}
\label{sec:evospex}

\begin{figure}[t!]
\begin{center}
\includegraphics[width=8cm]{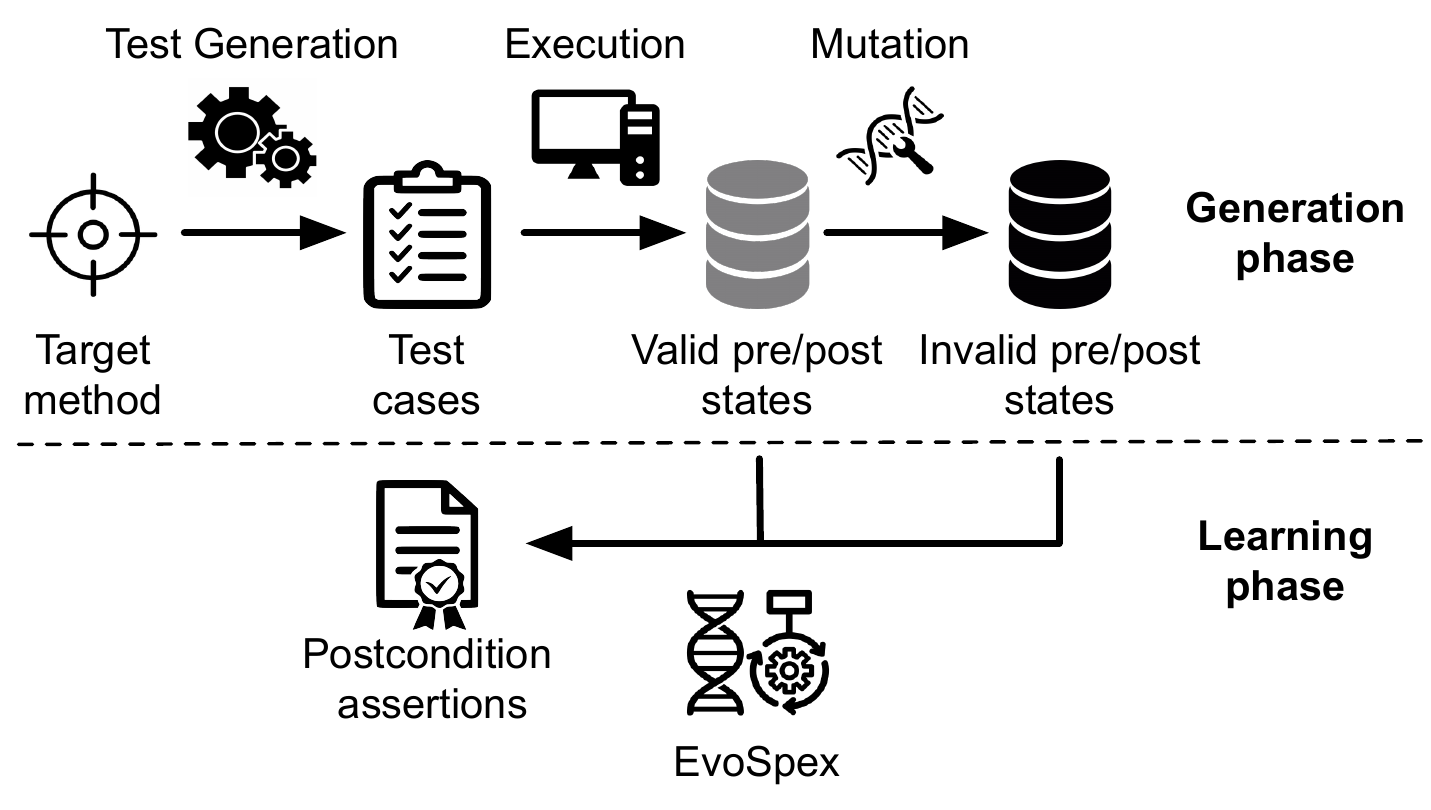}
\end{center}
\caption{An overview of the proposed approach}
\label{overview-approach}
\end{figure}

We now present the details of our technique for inferring method postconditions. An overview is shown in Fig.~\ref{overview-approach}. The technique is composed of two main phases: state generation and learning. During state generation, we produce pre/post program state pairs which are later on used in the learning phase to guide the search for suitable postcondition assertions. Two kinds of state pairs are generated: \emph{valid} ones, which capture actual method behaviors that candidate assertions should satisfy; and \emph{invalid} ones, which \emph{attempt} to capture incorrect behaviors (pre/post pairs that do not correspond to the current method behavior), that candidate assertions should not satisfy. Valid pre/post pairs are generated by executing the target method, using a test generation technique; clearly, these pairs correspond to the behavior of the method, as they were generated from its execution. Invalid pre/post pairs, on the other hand, are generated by \emph{mutating} valid pairs, going out of the set of valid pairs; contrary to the case of valid pairs, it is not guaranteed that our invalid pairs are indeed incorrect method behaviors, i.e., that they represent behaviors that are \emph{not} exhibited by the method. This may clearly affect the precision of the obtained assertions, since the algorithm would be guided to avoid some allegedly invalid behaviors which are actually valid. In these situations, the obtained assertions would be stronger than necessary, leading to a higher number of false positives, when evaluating assertion quality. We consider this issue in the design of our technique, in the following way. Firstly, the effectiveness in generating truly invalid pairs depends on the exhaustiveness of the set of valid pairs: the more exhaustive the set of valid pairs, the greater the chances that mutating out of this set leads to a truly incorrect method behavior. Secondly, as the soundness of the mechanism for invalid state pair generation cannot be guaranteed, one may risk favoring incorrect assertions based on wrong invalid state assumptions. The former motivates the use of a bounded-exhaustive test generation approach for valid state pairs. The latter drives an asymmetric treatment of valid and invalid state pairs in the fitness function, that gives the reliable information provided by valid pairs a greater relevance. We further describe in this section how we handle these issues, as well as other details of the genetic algorithm, and in the next section we evaluate the technique, including an evaluation of assertion precision.


\subsection{Generation of Valid/Invalid Method Executions}

The learning phase of our algorithm depends on a set of valid/invalid method executions, which guide the search for postcondition assertions. This is an important part of our algorithm. The overall process starts by generating runs of the target method $m$, collecting the pre/post states $\langle s, s' \rangle$ of each execution; these are the \emph{valid} execution pairs $V$. In order to generate \emph{invalid} execution pairs $I$, valid pairs are \emph{mutated}: for a valid pair $\langle s, s' \rangle$, we mutate $s'$ into $s''$, and check that $\langle s, s'' \rangle$ does not belong to $V$, to consider it part of $I$. Of course, the mutated pre/post pair may actually correspond to a valid execution of $m$ that we had not generated in $V$. The effectiveness of this latter approach depends on how thorough $V$ is (although we may still generate ``unseen'' valid execution pairs via mutation), motivating a bounded exhaustive approach for generating valid execution pairs.

The mechanism for generating valid execution pairs works as follows. Let $C, C_{1}, \dots, C_{n}$ be classes, and $m$, the target method, a method in $C$ with parameters of types $C_{1}, \dots, C_{n}$. The initial states for the execution of $m$ will be tuples $\langle o_C, o_{C_1}, \dots, o_{C_n} \rangle$ of objects of types $C, C_{1}, \dots, C_{n}$, respectively. We build the objects to form these tuples, for each class, bounded exhaustively, in the following way. Let $C_i$ be a class, and methods $b_1, \dots, b_l$ a set of \emph{builders} for $C_i$, i.e., a set of manually identified methods that can be used to create objects of class $C_i$. For instance, for a set collection, builders would include constructors and insertion routines. Given a bound $k$ (maximum length for method sequences), we build a set of objects of class $C_i$ using a variant of Randoop \cite{Pacheco+2007}. Randoop randomly generates sequences of methods of $C_i$'s API, of increasing length, by iterating a process in which previously produced traces are randomly selected, together with a method, to generate a new trace that calls this method. Our variant incorporates two main modifications to this process:
\begin{itemize}

\item The random selection of a method to extend a previously produced trace $t$ (test case), implemented in \cite{Pacheco+2007}, is replaced by a mechanism to systematically select \emph{all} methods in $b_1, \dots, b_l$, leading to $l$ different extensions of $t$. This is applied until the bound $k$ is reached. 

\item A state matching mechanism is implemented, to reduce the number of method combinations: when a newly produced trace leads to an object that matches a previously collected one, the trace (and the object) are discarded. The state matching approach borrows the canonical object representation put forward in \cite{DBLP:conf/icse/PonzioBBSAF18}. 

\end{itemize}
Besides the bound $k$ on trace length, the state matching mechanism also requires a maximum number of objects per type, and a range for primitive types (e.g., 0..k-1 for integers). This is a $k$-based scope, as defined in finitization procedures in \cite{Boyapati+2002} (a standard issue in bounded exhaustive generation). Using the above mechanism, we build the tuples of initial states, to execute $m$. We execute $m$ in each of these tuples, and collect the corresponding post-states, building in this way the set $V$ of \emph{valid} pre states and corresponding post states for $m$. 

The mutations applied to produce the ``invalid'' pre/post state set $I$, take a valid execution pair $\langle s, s' \rangle$, and create $\langle s, s'' \rangle$, where $s''$ mutates $s'$ by selecting a random field in the receiving object or return value (the constituents of $s'$), and replacing the value by a randomly generated value of the corresponding type within the above mentioned scope. We check that the resulting pair is not in $V$ before including it into the invalid state pair set $I$. 

\subsection{Chromosomes representing Candidate Postconditions}

Our representation of candidate assertions is based on the encoding used in \cite{Molina+2016}, where chromosomes represent conjunctions of assertions (each gene in a chromosome represents an assertion). That is, given a chromosome $c$, the candidate postcondition $\varphi_{c}$ represented by $c$ is defined as follows:
\begin{displaymath}
c = \langle g_{1} , g_{2} , \ldots , g_{n} \rangle \Rightarrow \varphi_{c} = g_{1} \land g_{2} \land ... \land g_{n} 
\end{displaymath}
As opposed to what is most common in genetic algorithms, chromosomes have varying lengths in this representation (up to a maximum chromosome length), and gene positions are disregarded by the genetic operators (see below), due to the associativity and commutativity of the conjunction. Genes need to encode complex assertions. Below we describe how genes are built, mutated and combined.  

\subsection{Initial Population}

\begin{figure}[t!]
\begin{center}
\includegraphics[width=8cm]{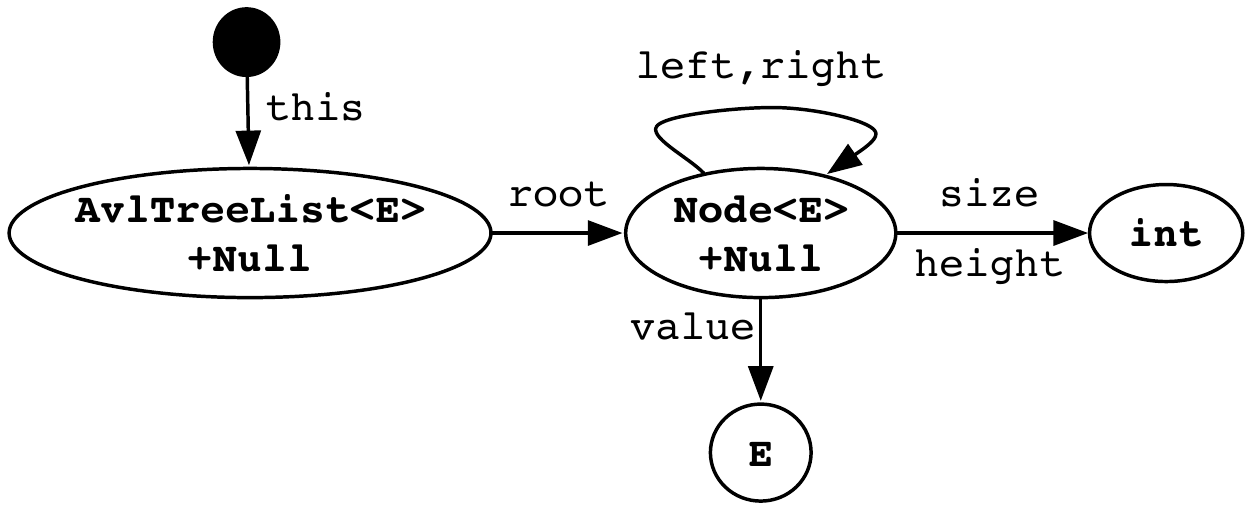}
\end{center}
\caption{Type graph for AvlTreeList example.}
\label{avl-type-graph}
\end{figure}

Let us describe how we build the initial population, to start our genetic algorithm. In order to create individuals representing ``meaningful'' postconditions, i.e., assertions stating properties of objects that are reachable at the end of the method executions, we take into account typing information, as in \cite{Molina+2016}. We consider a \textit{type graph} built automatically from the class under analysis: nodes represent types, and each field $f$ of type $B$ in class $A$ will produce an arc in the graph going from the node representing $A$ to the node representing $B$. For example, if we consider the \texttt{AvlTreeList} class in Figure~\ref{treelist}, the corresponding type graph would be the one shown in Figure~\ref{avl-type-graph}. It is straightforward to see that by traversing the graph, typed expressions can be built, using the fields of the object from which the method was executed. Some examples are \texttt{this.root}, \texttt{this.root.left}, \texttt{this.root.size}, \texttt{this.root.value}, and so on. Moreover, from loops in the graph, expressions denoting sets, such as \texttt{this.root.*left} (the set of nodes reachable from \texttt{this.root} via \texttt{left} traversals only), \texttt{this.root.*right} and \texttt{this.root.*(left+right)}, can be created (as explained earlier, we are using \texttt{*} for reflexive-transitive closure, as in \cite{Jackson2006}). Size one chromosomes are created using expressions denoting a single value, evaluating these on a randomly selected subset of the valid (resp. invalid) method executions, in the following way: if the result of evaluating an expression \texttt{expr} in a valid (resp. invalid) tuple $t$ returns a value $v$, then we create the individual $\langle$\texttt{expr == v}$\rangle$ (resp. $\langle$\texttt{expr != v}$\rangle$). 

In addition to these basic individuals we also create chromosomes containing comparisons of random expressions of the same type (e.g., \texttt{this.root == this.root.right}), chromosomes with quantified formulas considering expressions denoting sets (e.g., \texttt{all n: this.root.*(left+right) - null : n == n.right}), and individuals comparing integer expressions with the cardinality of expressions denoting sets (e.g., \texttt{this.root.height == \#(this.root.*right)}). Finally, since the method under analysis may have a return value or a set of arguments, we also include, in the set of initial candidates, expressions comparing them against expressions of the same type (e.g., \texttt{result < this.f}). The expressions used to compare with the result variable or the arguments, as well as the operators, are randomly chosen.

Notice that \emph{all} our initial chromosomes are size 1 chromosomes. The main reason for this design choice is to allow the genetic algorithm to progressively produce complex candidate postconditions by means of the genetic operators, that we define later on in this section. While this size-one chromosomes for the initial population is non-standard in genetic algorithms, in our case it helps the algorithm to more quickly converge to better fitted individuals. The replication package site  \cite{evospex-site} contains the results of comparing the effectiveness of our size-one chromosomes in the initial population, with standard size-$N$ chromosomes (we do not include the comparison here due to space restrictions). 

\subsection{Fitness Function}

Our fitness function assesses how good a candidate postcondition is, by distinguishing between the set $V$ of valid executions and the set $I$ of invalid executions. To do so, before computing the fitness value of a given candidate $c$, we obtain the postcondition $\varphi_{c}$ that $c$ represents, and then compute the sets $P$ and $N$ of positive and negative counterexamples, respectively. These sets are defined as follows:
\begin{displaymath}
    \begin{array}{cc}    
        P = \lbrace v \in V  | \neg \varphi_{c}(v) \rbrace &N = \lbrace i \in I  | \varphi_{c}(v) \rbrace  \\
    \end{array}
\end{displaymath}
where $\varphi_{c}$ is the postcondition represented by the candidate $c$. Basically, the sets $P$ and $N$ contain those executions for which the postcondition $\varphi_{c}$ does not behave correctly. Recall that, as opposed to the case of $V$ which reliably represents actual execution information of $m$, the set $I$ may contain mutated executions that are considered ``invalid'', but correspond to actual executions of $m$. This motivates a definition of our fitness function that does not treat $P$ and $N$ symmetrically. The fitness function $f(c)$ is computed as follows:
\begin{displaymath}
\begin{cases} 
	\#P > 0 \rightarrow (\textsf{MAX} - \#P - \#I)+\left( \frac{1}{l_{c} + comp_{c}} \right) + \frac{mca_{c}}{l_{c}} \\
	\#P = 0 \rightarrow (\textsf{MAX} - \#N)+\left( \frac{1}{l_{c} + comp_{c}} \right) + \frac{mca_{c}}{l_{c}}
\end{cases}
\end{displaymath}
This case-based definition aims at considering the negative counterexamples only when no positive counterexamples are obtained. In fact, for arbitrary candidates $c_1$ and $c_2$, if $c_1$ has no positive counterexample and $c_2$ has some positive counterexamples, then $f$ is guaranteed to produce worse fitness values for $c_2$, no matter how many negative counterexamples these candidates have. The rationale here is to make the reliable positive-counterexample information more relevant. 

The definition of the fitness function has three parts. The first term reflects the most important aspect: to minimize the number of counterexamples. The fact that when the candidate postcondition $\varphi_{c}$ has positive counterexamples, i.e., it is falsified by a correct method execution, the whole set $I$ of invalid executions is considered as counterexamples too, is what guarantees our above observation regarding the prioritization of candidates with no positive counterexamples. More precisely, the first term of the function subtracts $\#I$ when $\#P > 0$, to ensure that the fitness value of such individual is lower that the fitness value of any other individual that only has negative counterexamples. The second term of the fitness function acts as a penalty regarding two aspects: the candidate length $l_{c}$ and the candidate ``complexity'' $comp_{c}$. The candidate length is simply the number of conjuncts in the assertion, and it is considered in order to guide the algorithm towards producing smaller assertions. The candidate complexity is the sum of each conjunct complexity. Intuitively, the complexity of an equality between two integer fields is lower than the complexity of an equality between an integer field with a set cardinality, and both of these are lower than the complexity of a quantified formula, and so on. The last term of the function acts as a reward favoring those candidates with a greater number of ``method component assertions'' $mca_{c}$, i.e., with a high number of conjuncts of the candidate postcondition that represent properties regarding the parameters, the result, or a relation between initial and final object states. As described, the penalty related to the candidate length and complexity as well as the reward prioritizing the method component assertions just contribute a fraction to the fitness value, since we want the algorithm to focus on individuals whose number of counterexamples is approaching zero.  

\subsection{Genetic Operators}
\label{gen-op}

During evolution, the genetic operators allow the algorithm to explore the search space of candidate solutions, by performing certain operations that produce individuals with new characteristics as well as combinations of existing ones. In particular, our algorithm implements two well known genetic operators, namely the \textit{mutation} and \textit{crossover} operators. Some of these genetic operators were inspired by similar ones introduced in \cite{Molina+2016}, while others are novel. Also, a custom \textit{selection} operator was implemented, to keep in the population those candidates that are more suitable to be part of the real post condition.

Each chromosome gene is selected for mutation with a probability of $0.3$, and the operation can perform a variety of modifications depending on the shape of the selected gene expression. From a general point of view, the set of considered mutations are the following:

\textbf{Gene deletion:} it can be applied to any gene and simply removes the gene expression from the chromosome.

\textbf{Negation:} it negates the gene expression and is applied to any gene except quantified assertions.

\textbf{Numeric addition/subtraction:} it is only applied to genes that compare two expressions evaluating to a number, and it adds/subtracts a randomly selected numeric expression to the right-hand side of the comparison. 

\textbf{Expression replacement:} it applies to any gene, and it replaces some part of the gene with a randomly selected expression of the same type. 

\textbf{Expression extension:} it can be applied to any gene that involves a navigational expression, and it extends this expression with a new field, for example replacing \texttt{this.root} by \texttt{this.root.left}.

\textbf{Operator replacement:} it replaces an operator by an alternative one. The operators vary depending on the current gene expression. For instance, for relational equalities, the possible operators are $\lbrace ==, !=\rbrace$; for numeric comparisons, the operators are $\lbrace ==, !=, <, >, <=, >=\rbrace$; and for quantified expressions, the quantifiers are $\lbrace all, some\rbrace$. 

To produce combinations of individuals, we use a crossover rate of $0.35$. Given two randomly selected chromosomes $c_{1}$ and $c_{2}$, our \textit{crossover} operator simply produces a new individual that contains the union of the genes of $c_{1}$ and $c_{2}$, and thus represents the candidate postcondition $\varphi_{c1} \land \varphi_{c2}$. An important detail in our \textit{crossover} operator is that before selecting individuals for combination, we filter the population, keeping individuals which only have negative counterexamples, i.e., that represent formulas that are consistent with all valid method executions. The main reason for this policy is that we want the algorithm to join chromosomes that are already consistent with the valid method executions. 

Finally, to keep in the population the \textit{best} candidates of each generation, our \textit{selection} operator is defined as follows: given a number $n$ to be used as constant population size, our operator first sorts all the candidate postconditions in decreasing order, and then the candidates to be moved to the next generation are the first $n/2$ individuals plus the best $n/2$ unary non-valid individuals, i.e., size 1 chromosomes whose only gene is a formula that still has positive counterexamples. Additionally, our operator keeps all the unary valid candidates, that is, those that only have negative counterexamples. This last policy in our selection operator allows us to keep in the population all the discovered valid properties that the algorithm can use in future crossover operations.

\section{Evaluation}
\label{evaluation}

\newcommand\foobar[2][]{\lstinline[#1]{#2}}

\begin{center}
\begin{table*}[ht!]
\caption{Postconditions inferred by EvoSpex and Daikon after removing False Positives}
\scalebox{0.82}{
{
  \begin{tabular}{|p{3.8cm}||p{9cm}||p{7.5cm}|}
\hline

\centering{\textbf{Method}} & \multicolumn{1}{c||}{\centering{\textbf{EvoSpex}}} & \multicolumn{1}{c|}{\centering{\textbf{Daikon}}} \\

\hline
\multicolumn{3}{|>{\columncolor{black!5}}c|}{\textbf{jiprof - com.mentorgen.tools.profile.runtime.ClassAllocation} } \\
\hline
\foobar[language=java]{getAllocCount() : int} 
&
\foobar[language=java]{result == this._count}
&
\foobar[language=java]{this._count == result \&\& result == old(this._count)} 
\\
\hline
\foobar[language=java]{incAllocCount() : void} 
&
\foobar[language=java]{this._count == 1 + old(this_count)}
&
\foobar[language=java]{this._count >= 1 && this._count - old(this_count) - 1 == 0} 
\\
\multicolumn{3}{|>{\columncolor{black!5}}c|}{\textbf{jmca - com.soops.CEN4010.JMCA.JParser.SimpleNode} } \\
\hline
\foobar[language=java]{jjtSetParent(Node n) : void}
& 
\foobar[language=java]{n == this.parent}  
&
\foobar[language=java]{this.parent == old(n) \&\&} \newline
\foobar[language=java]{this.children == old(this.children) \&\&} \newline
\foobar[language=java]{this.id == old(this.id) \&\&} \newline
\foobar[language=java]{this.parser == old(this.parser) \&\&} \newline   
\foobar[language=java]{this.identifiers == old(this.identifiers)}
\\ 
\hline
\multicolumn{3}{|>{\columncolor{black!5}}c|}{\textbf{bpmail - ch.bluepenguin.email.client.service.impl.EmailFacadeState} } \\
\hline
\foobar[language=java]{setState(Integer ID, boolean dirtyFlag): void}
& 
\foobar[language=java]{ID in this.states.keySet()}
&
\foobar[language=java]{this.states == old(this.states)}
\\
\hline
\multicolumn{3}{|>{\columncolor{black!5}}c|}{\textbf{byuic - com.yahoo.platform.yui.compressor.JavaScriptIdentifier} } \\
\hline
\foobar[language=java]{preventMunging() : void}
& 
\foobar[language=java]{this.mungedValue == old(this.mungedValue) \&\&} \newline
\foobar[language=java]{this.refCount == old(this.refcount) \&\&} \newline
\foobar[language=java]{all n : this.declaredScope.*parentScope: n !in n.^parentScope}
&
\foobar[language=java]{this.mungedValue == old(this.mungedValue) \&\&} \newline
\foobar[language=java]{this.recCount == old(this.refcount) \&\&} \newline 
\foobar[language=java]{this.declaredScope == old(this.declaredScope) \&\&} \newline
\foobar[language=java]{this.markedForMunging == false}
\\ 
\hline
\multicolumn{3}{|>{\columncolor{black!5}}c|}{\textbf{dom4j - org.dom4j.tree.LazyList} } \\
\foobar[language=java]{add(E element): boolean}
&
\foobar[language=java]{old(this.size) == this.size - 1 \&\&} \newline
\foobar[language=java]{result == true \&\&} \newline
\foobar[language=java]{element in this.header.*next.element}
&
\foobar[language=java]{this.header == old(this.header) \&\&} \newline
\foobar[language=java]{this.size >= 1 \&\&} \newline
\foobar[language=java]{result == true \&\&} \newline
\foobar[language=java]{this.size - old(this.size) - 1 == 0}
\\ 
\hline
\end{tabular}}
}
\label{ga-vs-daikon}
\end{table*}
\end{center}

To evaluate our technique, we performed experiments focused on the following research questions:

\begin{itemize}

\item[RQ1] \emph{Do the oracles learned by EvoSpex have any deficiency compared to oracles produced by related tools?}

\item[RQ2] \emph{Are the assertions produced by the algorithm close to manually written contracts?}

\end{itemize}

To evaluate RQ1, we need to consider programs (in our case, Java programs) for which to infer method specifications. As mentioned earlier in the paper, and as it is clear from our candidate assertion state space and evolution operators, we target classes and methods with reference-based implementations, in particular classes where the corresponding internal representation has strong (implicit) invariants. As a source for our benchmark, we considered SF110\footnote{https://www.evosuite.org/experimental-data/sf110/} (originally used in \cite{DBLP:journals/tosem/FraserA14}), a collection of 110 Java projects (100 random projects, plus the 10 most popular ones according to SourceForge), that covers a wide variety of software, representative of open source development. Our process of assessing postcondition assertions makes use of the OASIs tool \cite{DBLP:conf/issta/JahangirovaCHT16}, essentially, to evaluate the quality of a postcondition assertion in terms of its associated number of false positives and false negatives. The process of computing this number requires a manual process (as described in \cite{DBLP:conf/issta/JahangirovaCHT16}, to compute the false negatives one first needs to get rid of the false positives, which implies having to manually refine the produced postconditions every time OASIs reports the presence of a false positive). Therefore, we are unable to consider the whole 110 projects in the benchmark. We randomly selected 16 projects, skipping cases in our selection that have a clear dependency on the environment (our technique involves automated test generation, and environment dependencies seriously affect these tools). The 16 projects can be found in Table~\ref{effectiveness-table-sf110}. For each case study, we selected various methods with different behaviors for analysis, manually defined a set of builders, and then generated the corresponding valid and invalid method executions with a relatively small scope (3 for all cases). Then, we executed our tool in the following way: for each method $m$ selected for analysis, we executed the genetic algorithm to produce a postcondition for $m$ until it reached 30 generations or a 10 minutes timeout was fulfilled. We repeated this execution 10 times, and then selected the postcondition assertion that repeated the most number of times, from the 10 produced by the algorithm. Additionally, in order to compare our tool with related approaches, we executed Daikon to infer post conditions for each method $m$. It is important to remark that the test suites that we fed Daikon with to produce postconditions for the methods under analysis, were exactly the same test suites that were used to generate the valid method executions in our technique (our valid bounded exhaustive suites). Both our tool and Daikon can produce assertions leading to false positives (see Section 2 for a comment on this issue), as well as redundant assertions. 

The results of this experiment are shown in Tables~\ref{ga-vs-daikon} and \ref{effectiveness-table-sf110}. Table~\ref{ga-vs-daikon} presents the postconditions generated by the tools, after removing the false positives and the redundant assertions, with the aim of giving a clear glance of the complexity of the assertions that the techniques are able to generate. We considered these assertions, as the ones produced by the two techniques. We then measured the quality of the corresponding assertions by automatically computing false positives and false negatives, using the OASIs \cite{DBLP:conf/issta/JahangirovaCHT16} tool. Table~\ref{effectiveness-table-sf110} shows the results of this quality assessment. Specifically, for each case study, we report in Table~\ref{effectiveness-table-sf110}: \emph{(i)} lines of code (LoC) of the evaluated project; \emph{(ii)} number of analyzed methods from the corresponding project; \emph{(iii)} number of assertions produced as part of the postconditions; \emph{(iv)} amount and percentage of false positives present in all generated assertions; and \emph{(v)} number of methods for which false negatives were detected. Notice that, as proposed in \cite{DBLP:conf/issta/JahangirovaCHT16}, false negatives detection is performed once all the false positives have been removed from the postcondition (hence the manual task that made us consider a subset of SF110). For both false positives detection and false negatives detection, we executed OASIs with a timeout of one minute. Problems with OASIs prevented us from reporting the number of false negatives for each method and case study; more precisely, when the tool reported the existence of false negatives, in some cases it was unable to produce the witnessing counterexamples (test cases), preventing us from measuring the number of false negatives identified by the tool in these cases. This issue was discussed with the developers of the tool. We therefore inform the number of methods for which OASIs reported the existence of false negatives, rather than the number of false negatives found, as this information was not reliably produced by the tool for all cases. For instance, for project \texttt{imsmart}, out of the 3 methods analyzed, OASIs found one of the corresponding assertions discovered by Daikon to have false negatives, and one of the assertions discovered by EvoSpex to have a false negative, too.

The evaluation of RQ2 requires having classes with methods featuring manually written contracts. Moreover, as discussed in Section 2, assertions for run-time checking are typically weak, efficiently checkable assertions, that weakly capture the semantics of the corresponding classes and methods \cite{Meyer1997,DBLP:conf/vstte/PolikarpovaFM10}. In order to compare with strong contracts, we took:
\begin{itemize}
    \item  A set of case studies with contracts written for the verification of object oriented programs. More precisely, these programs are written in Eiffel \cite{Meyer1998}, and the accompanying contracts were used for verification using the AutoProof tool \cite{TFNP-TACAS15}, a verifier for object-oriented programs written in the Eiffel programming language, for Eiffel programs. 
    \item A set of case studies for which a data representation and method implementations are automatically synthesized from a higher-level specification. More precisely, the synthesized implementations are taken from \cite{DBLP:conf/icse/LoncaricET18}, are generated by the Cozy tool, and are guaranteed to be correct with respect to higher level specifications, which serve as method contracts.
\end{itemize}

From \cite{TFNP-TACAS15}, we specifically considered various methods and their corresponding postconditions, from the following cases: 
\begin{itemize}
    \item  \textbf{Composite}: A tree with a consistency constraint between parent and children nodes. Each node stores a collection of its children and a reference to its parent; the client is allowed to modify any intermediate node. A value in each node should be the maximum of all children's values. 
    \item \textbf{DoublyLinkedListNode}: Node in a (circular) doubly-linked list with a structural invariant enforcing that its left and right links are consistent with its neighbors. 
    \item \textbf{Map$<$K,V$>$}: Generic Map abstract datatype implementation, based on two lists that contain the keys and values, and with operations that perform linear searches on the lists.
    \item \textbf{RingBugger$<$G$>$}: Bounded queue implemented over a circular array. 
\end{itemize}
Since our tool is for Java, and these implementations are in Eiffel, we had to manually translate the whole classes into Java, for analysis with our tool (this also prevented us from considering more sophisticated case studies in this evaluation). While the translation was manual, we made an effort in making it systematic, preserving the structure of the original code, and taking into account the semantics of references (e.g., expanded types in Eiffel), array indexing in Eiffel vs. Java, etc., using as a guideline the J2Eif work \cite{DBLP:conf/tools/TrudelOFN11}. Eiffel also differs from Java in other important aspects that did not affect the translation (e.g., inheritance, visibility of features, etc.). While we did not formally verify our translation, it was code-reviewed independently by co-authors of the paper.

From \cite{DBLP:conf/icse/LoncaricET18}, we considered several high-level specifications and their corresponding synthesized Java implementations: 
\begin{itemize}
    \item \textbf{Polyupdate}, a bag of elements that keeps track of the sum of its positive elements. 
    \item \textbf{Structure}, a simple class encapsulating a function and caching a parameter. 
    \item \textbf{ListComp02}, a structure composed of two collections of different elements, and operations that combine elements of the collections. 
    \item \textbf{MinFinder}, a bag of elements with a min operation. 
    \item \textbf{MaxBag}, a set of elements, with a max operation.
\end{itemize}

In order to infer postconditions for methods in these classes, we first generated valid and invalid method executions, as described earlier in this paper, for each of the target methods using a scope of 4. Then, we executed our algorithm using the same configuration described for RQ1 (30 generations with a 10-minute timeout). Again, we repeated the execution 10 times and selected the most frequently obtained postcondition. Notice that our approach is not using the real contracts already accompanying the target methods. We fully ignore these in the inference approach, and only consider the methods source code, both for the generation of valid/invalid method executions, and for the actual evolutionary inference. A similar procedure was followed for the Cozy case studies. We computed postconditions for the Java implementations, and contrasted them with those in the original high-level specifications, from which the Java implementations were derived. 

The results of this experiment are shown in Tables~\ref{eiffel-contracts} and \ref{cozy-collections}. In Table~\ref{eiffel-contracts}, for each of the target methods, the column Eiffel Contracts lists the properties that are present in the original postcondition (expressed as text, for easier reference). In Table~\ref{cozy-collections}, the original postcondition is described in column High-level spec in terms of the abstract state declared in the specification. In both tables, the column EvoSpex indicates which of the corresponding assertions in the original contract, our evolutionary algorithm was able to infer. Finally, Table~\ref{summary-rq2} summarizes these results and also reports the number of \emph{invalid} assertions synthesized as part of the inferred specifications for each subject in Eiffel and Cozy case studies. 

\begin{table}[ht!]
\caption{Measuring the quality of postconditions inferred by Daikon and EvoSpex, using OASIs.}
\label{effectiveness-table-sf110}
\begin{center}
\begin{scriptsize}
\renewcommand{\arraystretch}{1.2} 
\setlength{\tabcolsep}{0.4em}
\begin{tabular}{llllr|rr|c}
\hline
\textbf{Project} & \textbf{LOCs} & \textbf{Methods} & \textbf{Technique} & \textbf{\#Assertions} &\multicolumn{2}{c|}{\textbf{FPs}} & \textbf{FNs} \\
\hline
& & & &  & Total & \% & \\
\hline
imsmart & 1407 & 3 & Daikon & 21 & 2 & 9.52 & 1 \\
& & & EvoSpex & 4 & 1 & 25 & 1 \\
\hline
beanbin & 4784 & 5 & Daikon & 35 & 5 & 14.29 & 0 \\
& & & EvoSpex & 7 & 0 & 0 & 0 \\
\hline
byuic & 7699 & 7 & Daikon & 165 & 21 & 12.73 & 4 \\
& & & EvoSpex & 36 & 4 & 11.11 & 2 \\
\hline
geo-google & 20974 & 7 & Daikon & 93 & 30 & 32.26 & 0 \\
& & & EvoSpex & 10 & 3 & 30 & 4 \\
\hline
templateit & 3315 & 7 & Daikon & 37 & 4 & 10.81 & 3\\
& & & EvoSpex & 20 & 0 & 0 & 2 \\
\hline
water-simulator & 9931 & 9 & Daikon & 39 & 3 & 7.69 & 9 \\
& & & EvoSpex & 18 & 3 & 16.67 & 9 \\
\hline
dsachat & 5546 & 9 & Daikon & 138 & 15 & 10.87 & 3 \\
& & & EvoSpex & 18 & 2 & 11.11 & 2 \\
\hline
jmca & 16891 & 9 & Daikon & 205 & 26 & 12.68 & 0 \\
& & & EvoSpex & 25 & 1 & 4 & 3 \\
\hline
jni-inchi & 3100 & 10 & Daikon & 122 & 12 & 9.84 & 2 \\
& & & EvoSpex & 50 & 1 & 2 & 4 \\
\hline
bpmail & 2765 & 11 & Daikon & 46 & 6 & 13.04 & 8 \\
& & & EvoSpex & 17 & 0 & 0 & 7 \\
\hline
dom4j & 42198 & 18 & Daikon & 166 & 27 & 16.27 & 7 \\
& & & EvoSpex & 25 & 2 & 8 & 10 \\
\hline
jdbacl & 28618 & 19 & Daikon & 115 & 17 & 14.78 & 10 \\
& & & EvoSpex & 80 & 3 & 3.75 & 8 \\
\hline
jiprof & 26296 & 20 & Daikon & 352 & 81 & 23.01 & 20 \\
& & & EvoSpex & 35 & 4 & 11.43 & 19 \\
\hline
summa & 119963 & 21 & Daikon & 273 & 67 & 24.54 & 6 \\
& & & EvoSpex & 62 & 5 & 8.06 & 5 \\
\hline
corina & 78144 & 22 & Daikon & 155 & 13 & 8.39 & 17 \\
& & & EvoSpex & 55 & 1 & 1.82 & 17 \\
\hline
a4j & 6618 & 23 & Daikon & 257 & 59 & 22.96 & 9 \\
& & & EvoSpex & 60 & 5 & 8.33 & 5 \\
\hline
\hline
TOTAL &  & 200 & Daikon & 2219 & 388 & 17.49 & 99 \\
& & & EvoSpex & 522 & 35 & \bf{6.70} & \bf{98} \\
\hline
\end{tabular}
\end{scriptsize}
\end{center}
\end{table}

Our tool, all the case studies, and a description of how to reproduce the experiments presented in this section can be found in the site of the replication package of our approach \cite{evospex-site}. All the experiments were run on an Intel Core i7 3.2Ghz, with 16Gb of RAM, running GNU/Linux (Ubuntu 16.04).

\newcommand{\cmark}{\ding{51}}%
\newcommand{\xmark}{\ding{55}}%

\begin{table}[ht!]
\caption{Comparing manually written contracts (in Eiffel verified classes) with postconditions inferred by EvoSpex.}
\label{eiffel-contracts}
\centering
\setlength{\tabcolsep}{0.7em}
\renewcommand{\arraystretch}{1.2} 
\setlength\dashlinedash{1pt}
\scriptsize
\begin{tabular}{l|l|c}
\hline
\textbf{Method} & \textbf{Eiffel Contracts} & \textbf{EvoSpex} \\
\hline
\multicolumn{3}{c}{\cellcolor{Gray1}\textbf{Composite}}\\
add\_child(Composite c) : void & child added & \cmark  \\
& c value unchanged &  \\ 
& c children unchanged & \\
& ancestors unchanged & \cmark \\
\hline
\multicolumn{3}{c}{\cellcolor{Gray1}\textbf{DoublyLinkedListNode}}\\
insert\_right(DoublyLinkedListNode n) : void & n left set &  \cmark \\
& n right set &  \\
\cdashline{1-3}
remove() : void & singleton &  \cmark \\ 
& neighbors connected & \\ 
\hline
\multicolumn{3}{c}{\cellcolor{Gray1}\textbf{Map$<$K,V$>$}}\\
count() : int & result is size &  \cmark \\
\cdashline{1-3}
extend(K k,V v) : int & key set &  \cmark \\
& data set &  \cmark \\
& other keys unchanged &  \\
& other data unchanged &  \\
& result is index &  \\
\cdashline{1-3}
remove(K k) : int & key removed & \cmark \\
& other keys unchanged &  \\
& other data unchanged &  \\
& result is index &  \\
\hline
\multicolumn{3}{c}{\cellcolor{Gray1}\textbf{RingBuffer$<$G$>$}}\\
count() : int & result is size &  \\
\cdashline{1-3}
extend(G a\_value) & value added & \cmark \\
\cdashline{1-3}
item() : G & result is first elem &  \\
\cdashline{1-3}
remove() : void & first removed & \\
\cdashline{1-3}
wipe\_out() : void & is empty & \cmark \\
\hline
\end{tabular}
\end{table}

\begin{table}[ht!]
\caption{Inferring postconditions of synthesized collections.}
\label{cozy-collections}
\centering
\scriptsize
\setlength{\tabcolsep}{0.4em}
\renewcommand{\arraystretch}{1.2} 
\setlength\dashlinedash{1pt}
\begin{tabular}{ll|l|c}
\hline
\textbf{High-level state} & \textbf{Method} & \textbf{High-level spec} & \textbf{EvoSpex} \\
\hline
\multicolumn{4}{c}{\cellcolor{Gray1}\textbf{Polyupdate}}\\
x : Bag$<$Int$>$ & a(Integer y) : void & y added to x & \cmark \\
s : Int & & y added to s if positive &  \\ 
\cdashline{2-4}
& sm() : Integer & result is s + sum of x & \cmark \\  
\hline
\multicolumn{4}{c}{\cellcolor{Gray1}\textbf{Structure}}\\
x : Int & foo() : Integer & result is x + 1 & \cmark \\  
\cdashline{2-4}
& setX(Integer y) & x = y & \cmark \\  
\hline
\multicolumn{4}{c}{\cellcolor{Gray1}\textbf{ListComp02}}\\
Rs : Bag$<$R$>$ & insert\_r(R r) : void & r added to Rs & \cmark \\
\cdashline{2-4}
Ss : Bag$<$S$>$ & insert\_s(S s) : void & s added to Ss & \cmark \\
\cdashline{2-4}
& q() : Integer & result is the sum of Rs $\otimes$ Ss &  \\
\hline
\multicolumn{4}{c}{\cellcolor{Gray1}\textbf{MinFinder}}\\
xs : Bag$<$T$>$ & findmin() : T & result is min of xs & \cmark \\
\cdashline{2-4}
& chval(T x, int nv) : void & inner value of T is x \\
\hline
\multicolumn{4}{c}{\cellcolor{Gray1}\textbf{MaxBag}}\\
l : Set$<$Int$>$ & get\_max() : Integer & result is max of l & \cmark \\
\cdashline{2-4}
& add(Integer x) : void & x added to l & \cmark \\
\cdashline{2-4}
& remove(Integer x) : void & x removed from l & \\
\hline
\end{tabular}
\end{table}

\subsection{Assessment}

Let us briefly discuss the results of our evaluation. Regarding RQ1, the results show that our approach is able to generate postconditions containing more complex assertions than the ones produced by Daikon. This is mainly due to the fact that our technique focuses on evolving assertions targeting reference-based conditions in reference-based implementations, as opposed to Daikon whose expressions are comparatively simpler properties, that do not include complex structural constraints, membership checking, etc (with the exception of arrays and implementations of \texttt{java.util.List}, for which Daikon generates interesting structural properties). Furthermore, as Table~\ref{effectiveness-table-sf110} shows, a significant number of the assertions inferred by our technique are \emph{true positives}, i.e., assertions that hold for all valid post-states of the corresponding methods, for any scope. Of course, this check for true positives is in the end manual (we carefully analyzed how each of the evaluated methods operates, and inspected the obtained assertions after filtering out assertion conjuncts as per OASIs assessment); the oracle deficiency analysis performed by OASIs is inherently incomplete, we cannot guarantee the truth of the remaining assertions. 

\begin{table}[h!]
\caption{Summary of EvoSpex assertions on RQ2 subjects}
\label{summary-rq2}
\begin{center}
\renewcommand{\arraystretch}{1.2} 
\scriptsize
\setlength{\tabcolsep}{0.6em}
\begin{tabular}{lcc|cc}
\hline
\textbf{Subject} & \textbf{Methods} & \textbf{Manual Contracts} &\multicolumn{2}{c}{\textbf{Inferred Assertions}} \\
& & & Total & Invalid \\ 
\hline
\multicolumn{5}{c}{\cellcolor{Gray1}\textbf{Eiffel}} \\
Composite & 1 & 4 & 7 & 0 \\
DoublyLinkedListNode & 2 & 4 & 5 & 0 \\
Map$<$K,V$>$ & 3 & 10 & 10 & 0 \\
RingBuffer$<$G$>$ & 5 & 5 & 30 & 4 \\
\hline
\multicolumn{5}{c}{\cellcolor{Gray1}\textbf{Cozy}}\\
Polyupdate & 2 & 3 & 3 & 0 \\
Structure &  2 & 2 & 2 & 0 \\
ListComp02 & 3 & 3 & 6 & 0 \\
MinFinder & 2 & 2 & 3 & 0 \\
MaxBag & 3 & 3 & 33 &  5 \\
\hline
\end{tabular}
\end{center}
\end{table}

As shown in Table~\ref{effectiveness-table-sf110}, in most of the case studies (13 out of 16), the percentage of false positives that our tool generates, when considering the total amount of produced assertions, is less than that produced by Daikon. Thus, comparing it with Daikon, and solely based on false positives, our assertions are significantly more precise. In fact, in a total of 200 methods analyzed, our technique had a 6.7\% false positives, compared to the 17.49\% of Daikon (an order of magnitude improvement). Moreover, the relationship between the number of produced assertions (in total, 522 of EvoSpex vs. the 2219 produced by Daikon) and the identified presence of false negatives, shows that our technique produces overall assertions of similar strength, with significantly fewer constraints. Daikon seems to make a more heavy used of specifically observed values in the assertions, leading to assertions that, while true within the provided test suite cases, are violated when considering larger scopes. Our algorithm is guided both by valid and invalid pre/post method states, giving it an advantage over Daikon, and explore a state space of candidate assertions that are less affected by specific values observed in executions. Regarding false negatives, both Daikon and our technique lead to assertions for which OASIs is able to identify false negatives (with our technique having a small margin of advantage in this respect). The conclusion is clear: the assertions obtained with both tools are weaker than the ``strongest'' postcondition, thus letting ``pass'' undetected some mutations of the analyzed methods (cf. how OASIs identifies false negatives \cite{DBLP:conf/issta/JahangirovaCHT16}). Unfortunately, as discussed earlier, a problem with OASIs did not allow us to perform a more detailed comparison, based on the \emph{number} of identified false negatives in each case. Nevertheless, by inspecting the obtained postconditions, as shown in Table~\ref{ga-vs-daikon}, it is apparent that our technique produces stronger assertions. 

Regarding RQ2, the assertions that our technique can produce are close to those that may be defined by developers when manually specifying rich contracts. As Table~\ref{eiffel-contracts} shows, our algorithm generated at least one of the exact properties defined in the original assertions for the Eiffel methods in 8 out of 11 cases. Similarly, as Table~\ref{cozy-collections} indicates, in 9 out of 12 methods we correctly identified at least one property of the corresponding postcondition. This confirms that our technique is capable of generating assertions that are actually true positives and are scope-independent. In the case of the remaining properties that the algorithm was not able to infer, these are specific assertions regarding the arguments, complex properties over sets that are beyond the assertions that the algorithm may produce, such as the ``other keys unchanged'' in the Map.extend routine, or are relatively complex arithmetic constraints such as the ones present in Cozy's ListComp02 and the RingBuffer methods (notice that our assertions concentrate on properties of reference-based fields rather than sophisticated arithmetic assertions).  Regarding the precision of the generated specifications for Eiffel and Cozy case studies, it is also important to analyze if the tool produces invalid assertions. As Table~\ref{summary-rq2} shows, only 2 out of 9 subjects contained invalid assertions in the corresponding inferred postcondition, being all of them assertions that are true in the bounded scenarios from which they were computed, but are not if the scope is extended. These cases were the ones that involved a greater number of fields. Still the percentage of invalid postconditions for these cases were about 15\% or less (4 of 30 in one case, 5 of 33 in the other). Table~\ref{summary-rq2} also shows that, for most case studies, EvoSpex produces additional valid assertions, compared to the corresponding postcondition. Generally, these have to do with valid information that is not explicitly mentioned in the original postcondition. For instance, for Composite.addChild, EvoSpex produced a 7-conjunct postcondition, 2 of which are in the manual contract; the remaining 5 are either trivial (e.g., the list of children is not changed, the parent is not changed), or capture valid information not in the original ``ensure'' (e.g., acyclicity of the parent structure). For further details, we refer the reader to the replication package site \cite{evospex-site}, where all the assertions produced for each case study can be found. 

\section{Related Work}
\label{relatedwork}

Assertions can be exploited for a wide variety of activities in software development, notably program verification \cite{DBLP:conf/fmco/ChalinKLP05,DBLP:conf/sas/Fahndrich10} and bug finding \cite{DBLP:conf/tap/TillmannH08,DBLP:conf/sigsoft/LeitnerCOMF07}, but also including program comprehension, software evolution and maintenance \cite{DBLP:conf/icsm/SatpathySR04}, and specification improvement \cite{DBLP:conf/issta/JahangirovaCHT16,DBLP:conf/oopsla/SchillerE12}, among others. Thus, the problem of automatically inferring specifications from source code, and in general the problem of producing software oracles, has received increasing attention in the last few years \cite{Barr+2015}. Techniques for inferring specifications from source code, i.e., for \emph{deriving} oracles, include approaches based on program executions, such as those reported in \cite{Ernst+2007,Simons2007}, as well as some recent techniques based on machine learning \cite{Shahamiri+2011,DBLP:journals/fmsd/SharmaA16,DBLP:conf/icse/MolinaDPRAF19}. Compared to the execution based approaches, our technique is guided both by valid and invalid executions (actually, pre/post method states); compared to the machine learning approaches, our technique concentrates on method postconditions, and produces interpretable assertion in standard assertion languages, as opposed to assertions encoded into artificial neural networks and other machine learning models. A closely related technique is that proposed in \cite{Ernst+2007}, with which we compare in this paper. Tools for automated test generation, notably EvoSuite \cite{DBLP:conf/sigsoft/FraserA11} as well as Randoop \cite{Pacheco+2007} and some extensions \cite{DBLP:conf/issta/YatohSIH15}, can produce assertions accompanying the generated tests. However, these assertions are scenario-specific, i.e., they capture properties particular to the generated tests, as opposed to our postconditions that attempt to characterize general method behaviors. 

Our technique embeds a mechanism for test input generation, that follows a bounded exhaustive testing approach. As opposed to the previous mechanisms for generating bounded exhaustive suites, e.g., via tools like Korat \cite{Boyapati+2002} or TestEra \cite{Khalek+2011}, our technique generates bounded exhaustive suites from the program's API, rather than from an invariant specification. In this sense, our technique is more closely related to Randoop \cite{Pacheco+2007}, replacing the random method selection in building test traces, with a systematic generation of \emph{all} bounded method traces. The state matching mechanism we used in this paper is crucial in making this approach effective, but its discussion is beyond the scope of the paper. Besides producing valid method executions in the search of assertions, our technique also produces invalid program executions. The approach is based on mutating state. It is somehow related to the oracle assessment approach (for false negatives) implemented in the OASIs tool \cite{DBLP:conf/issta/JahangirovaCHT16}, although therein the authors mutate \emph{programs} (source code), as opposed to mutating \emph{state}. The idea of mutating state is used elsewhere, e.g., in \cite{DBLP:conf/icst/MalikSK11,DBLP:conf/icse/MolinaDPRAF19}.

\section{Conclusion}
\label{conclusion}

The oracle problem has become a very important problem in software engineering, and within this context, oracle derivation or inference is particularly challenging \cite{Barr+2015}. In this paper, we have proposed an evolutionary algorithm for oracle inference, in particular for inferring method assertions in the form of \emph{postconditions}. Our technique features various novel characteristics, including a mechanism for generating test inputs bounded exhaustively, from a component's API, and the definition of a genetic algorithm whose state space of candidate assertions includes rich constraints involving method parameters, return values, internal object states, and the relationship between pre and post method execution states. Our experimental evaluation shows that our tool is able to produce more accurate assertions (stronger contracts in the sense of \cite{DBLP:conf/icse/PolikarpovaF0WM13}, with the associated benefits described therein), with a total of 6.70\% of false positives, compared to the 17.49\% of false positives of related techniques, for a set of randomly selected methods from a benchmark of open source Java projects. Furthermore, our evaluation shows that our technique is able to infer an important part of rich program assertions, taken from a set of case studies involving contracts for program verification and synthesis. 

This work also opens several lines for future work. On one hand, our genetic algorithm uses a finite set of genetic operators, in particular the ones used for mutation; extending the set of operators and exploring new ones may be necessary to increase the scope of properties that the algorithm may produce, especially when dealing with more sophisticated programs. Fitness functions in genetic algoritms play a crucial role in the quality of the solutions; adapting the fitness function of our algorithm in order to prioritize general aspects of method postconditions may considerably improve our results. 
Our experiments were based on the use of a variant of random generation for the production of bounded exhaustive test suites. Using alternative test suite generation approaches such as fully random generation may allow us to produce different postconditions. The existence of false negatives for our produced postcondition assertions also opens lines of improvement for our inference mechanism.

\section*{Acknowledgments}

The authors would like to thank the anonymous reviewers for their helpful feedback, and the OASIs authors for their assistance in using the OASIs oracle assessment tool. 

This work was partially supported by ANPCyT PICT 2016-1384, 2017-1979 and 2017-2622. Facundo Molina's work is also supported by Microsoft Research, through a Latin America PhD Award. 

\newpage

\bibliographystyle{plain}
\bibliography{bibliography}

\end{document}